\documentstyle[12pt,preprint,aps,epsf]{revtex}

\tighten
\begin{document}
\draft

\title{Experimental Persistence Probability for Fluctuating Steps}
\author{D.B. Dougherty, I. Lyubinetsky and E.D. Williams \\}
\address{Materials Research Science and Engineering Center,\\
University of Maryland, College Park, MD 20742-4111}
\author{M. Constantin, C. Dasgupta and S. Das Sarma}
\address{Department of Physics, Condensed Matter Theory Center,\\
University of Maryland, College Park, MD 20742-4111}
\date{\today}
\maketitle

\begin{abstract}
The persistence behavior for fluctuating steps on the $Si(111)$ 
$(\sqrt3 \times \sqrt3)R30^{0} - Al$ surface was determined by analyzing time-dependent 
STM images for temperatures between 770 and 970K.  The measured persistence probability 
follows a power law decay with an exponent of $\theta=0.77 \pm 0.03$. This is consistent 
with the value of $\theta= 3/4$ predicted for attachment/detachment limited step kinetics. 
If the persistence analysis is carried out in terms of return to a fixed reference position, 
the measured persistence probability decays exponentially. Numerical studies of the Langevin 
equation used to model step motion corroborate the experimental observations.
\end{abstract}  
\narrowtext
\pagebreak

With increasing interest in exploiting the special properties of nanostructures, characterization of 
the general stability or "memory" over long times of a fluctuating structure is likely to be valuable \cite{1}. 
The idea of persistence that has been developed theoretically in recent years provides one way of addressing 
these issues \cite{1,2,3}. The persistence probability $p(t)$ is defined simply as the 
probability that a random variable never crosses a chosen reference level within the time interval $t$. 
It is related to the First-Passage probability \cite{4} and is not determined by a small number of moments 
of the random variable.  The ideas of 
persistence can be directly related to the stability of crystalline structures by realizing that structural 
fluctuations occur at steps (e.g. the edges of crystal planes), and that the model classes that have been used 
to describe these one-dimensional boundaries can be related to the same model classes that have been 
investigated in theoretical studies of persistence \cite{5}.

In the last decade, the unifying concepts of the continuum step model and lattice models of varying 
complexity have been applied with great success to experimental observations of fluctuating steps on 
both metallic \cite{6,7} and covalently bonded \cite{8,9} materials. 
The resulting fundamental understanding has led to quantitative success in kinetic descriptions 
of experimentally measured structural transformations on the nanometer to micron scale 
\cite{6,8,10}. However, in addition to the deterministic envelope of kinetic 
processes, the stochasticity of step fluctuations will become an increasingly important component of 
structural evolution at the nanoscale. The concepts of persistence are well matched to evaluating the 
impact of many interesting consequences of such stochasticity, such as the average time required for a 
structure to $first$ fluctuate into an unstable configuration, or for neighboring structures to $first$ 
fluctuate into contact. To develop such applications, it is necessary to demonstrate that the concepts 
of persistence are indeed applicable to step fluctuations. In this manuscript, we report an 
experimental demonstration of persistence phenomena on a solid surface, specifically for the 
temporal fluctuations of steps with moderately complex structure.

For crystalline structures, persistence can be defined in terms of the fluctuation 
of a step with respect to a chosen coordinate of interest. Numerous analytical and 
numerical studies on a wide variety of systems have indicated that the persistence 
probability has a universal scaling form 
\begin{equation}
\label{p}
p(t)  \propto t^{-\theta},
\end{equation}
\noindent where $\theta$, the persistence exponent, has a nontrivial value characteristic of the dynamics 
governing the evolution of the variable in question.  Systems ranging from the simple diffusion 
equation  \cite{2} to kinetic Ising models \cite{3} have been found to display the scaling form. 
Despite the number of interesting models studied theoretically, few experiments have been carried 
out to test the validity and utility of the persistence concept. Experiments on soap froth coarsening 
\cite{11}, liquid crystals \cite{12}, droplet condensation \cite{13}, and magnetization in 
spin polarized Xe gas \cite{14} appear to support the proposed universality of the persistence 
probability and agree with quantitative predictions of the persistence exponent.

We have applied persistence ideas to the case of fluctuating monatomic steps on a 
metal/semiconductor adsorption system $Si(111)$ $(\sqrt3 \times \sqrt3)R30^{0}-Al$ \cite{15,16}.
A fluctuation in the position of the step edge requires correlated motion of several atoms to effect 
the smallest unit of displacement, that of a single kink on the step edge. Because atomic processes 
in this system are more complex than in the lattice models often used in numerical studies, it 
provides a challenging test of universality.

Experiments were performed in UHV  (base pressure $\sim 6 \times 10^{-11}$ Torr) 
by depositing $0.25 \div 0.33$ ML of Al onto a clean $Si(111)-(7 \times 7)$ 
sample and imaging the resulting reconstructed surface at temperatures between 770 and 970 K with a 
Variable-Temperature STM (Omicron). More extensive experimental details have been published elsewhere 
\cite{17,18}. Images were obtained by scanning the STM tip repeatedly over a single point (e.g. fixed 
$y$ coordinate, scan in $x$-direction) on a step edge for time between 23 and 107 seconds. An example 
of one of these line-scan images is shown in Fig.1. The digitized step positions extracted from the 
images generate a function, $x(t)$, that can be used to study the statistics of fluctuating steps, 
for example the usual time correlation function:

\begin{equation}
\label{G}
G(t) = <\left( x(t) - x(0) \right)^{2}>
\end{equation}

\noindent as well as the persistence probability.

The time correlation function for equilibrium step fluctuations can be determined through a 
Langevin formalism, which has achieved great success over the past decade \cite{6,8,18}. In this 
phenomenological approach, the coarse-grained step position evolves in time via

\begin{equation}
\label{Langevin1}
\partial_{t} x(y,t)=F[x(y,t)]+\eta(y,t),
\end{equation}

\noindent where $F$ is the appropriate functional describing the relaxation of the step position and $\eta$ 
represents random noise due to thermal fluctuations.  The $y$ direction runs parallel to the 
average step edge so that $x(y,t)$ describes the step's perpendicular excursions from its average. 
For the two most commonly observed step fluctuation mechanisms, random attachment-detachment and 
diffusion along the step edge, $F$ is proportional to $(\nabla^{2})^{z/2}$, where the dynamic exponent 
$z$ equal to either $2$ or $4$ respectively \cite{6}. The statistical properties of the noise (volume 
conserving for $z = 4$ and white for $z = 2$) are chosen to satisfy relevant conservation laws and 
the fluctuation-dissipation theorem. For the two cases mentioned, the correlation function has the simple 
form,

\begin{equation}
\label{G2}
G(t) = c(T)~t^{1/z},
\end{equation}

\noindent where $c(T)$ contains the coarse-grained step stiffness and step mobility. A previous 
analysis of the time correlation function for this experimental system \cite{18} 
yielded an average exponent of $z = 2.17 \pm 0.09$ from a fit to Eq.(\ref{G2}) over the temperature range 
studied. Supporting evidence suggests that the rate-limiting step fluctuation mechanism is most 
likely random exchange of mass between the step edges and the intervening 
terraces, and that the system thus corresponds to the case with F in Eq.(\ref{G2}) 
proportional to $\nabla^{2}$ \cite{6,8,18}.  Extraction of $c(T)$ from the experimental data yielded 
time constants for elementary microscopic attachment/detachment events that vary from 
$1.2$ ms at $970 K$ to $260$ ms at $770 K$ \cite{18}.

The persistence probability is non-trivially related to the correlation function of the 
position-time data, and thus must be determined through a separate analysis. To do this, the 
experimental persistence probability was calculated by dividing the time axis from the image 
into intervals of length $t$ and then determining the fraction of those intervals for which 
the digitized step position never crossed a chosen reference. The persistence probability 
was determined from a number of images ($4 -14$) for each temperature, and the results for 
each temperature were averaged to obtain the final persistence probability.

The first-return question for each small time interval was investigated by choosing the reference as 
the initial step position in that specific interval. Figure 2 shows linear behavior on a double log 
scale, clearly indicating power law decay.  Between $770$ and $970 K$ the slopes in Fig.2 yield an 
average effective persistence exponent of $0.77 \pm 0.03$. This result agrees with computational 
predictions of a value of $0.75$ for the exponent for attachment-detachment limited step kinetics 
\cite{5} and for other systems in the same dynamic universality class. Thus, the two 
exponents, $z$ and $\theta$, computed independently from the same experimental data, lead to a 
consistent interpretation in terms of the model stochastic equation governing the rate-limiting 
step fluctuation mechanism.

The choice of the reference level used in the analysis above 
corresponds to a specific physical question. That is: Given the position $x$ of theof the step at some time zero, 
what is the probability that the step will not have returned to $t$? One can imagine a different physical question, 
relevant for equilibrated steps fluctuating about a known average position $x_{0}$. This question is: Given 
an arbitrary time zero, at which the step position may be any accessible value of $x$, what is the probability 
that the step position will not have returned to $x_{0}$ within time interval $t$? The second question 
corresponds to a persistence-like analysis with a fixed absorbing boundary.  Experimentally, the choice of 
reference level will be dictated by the specific purpose for which persistence is intended. In particular, 
first passage into some fixed region of a surface could easily influence the design and placement of small 
structures in nanometer scale electronic devices.

It was possible to analyze the data with a fixed return point because the surface under study was near equilibrium, 
and therefore the steps had well-defined average positions. Thus the average position computed from the entire STM 
time image was used as a fixed reference level. The resulting fixed-reference persistence probability was observed 
to decay exponentially with a time constant of $\sim 1.5$ seconds for all three temperatures investigated, as shown 
in Fig. 3. At short times, the data is better fit to a power law with an exponent of $\sim 0.4$ as shown in the 
inset. At long times for this choice of reference, very few segments of the image had avoided 
crossing the reference for the entire history of the evolution and the computed probability dropped 
precipitously to zero.

To model the experimental system, the $\nabla^{2}$ linear Langevin equation was solved numerically 
on small lattices (32 sites) using the discretization taken from Ref.\cite{5}. STM line-scan data 
was simulated by following one point of the lattice over the full time evolution and performing the 
calculations of persistence as described above. Figure 4 shows the results averaged over 100 independent 
numerical solutions for each choice of reference level. The behavior in both cases agrees with the 
experimental observations.  For the return to the initial configuration, a persistence exponent of $0.76$ is obtained 
as expected from Ref.\cite{5} and within the estimated experimental uncertainty of the present study. In addition, 
the rollover from power law to exponential decay associated with the average reference level occurs in the 
numerical integration as in the experimental data.

The clear power-law decay of the first return persistence probability for the fluctuations of steps 
on $Si(111)$ $(\sqrt3 \times \sqrt3)R30^{0}-Al$ demonstrates the feasibility of obtaining useful first-passage 
statistics from STM data. The fact that the same exponent is observed for three temperatures that correspond to 
more than two orders of magnitude difference in the underlying time constant of the physical system \cite{18} is a 
significant demonstration of the universal behavior predicted theoretically. Despite the complexity of the 
atomic-scale processes underlying the step fluctuations the value of the first return persistence exponent 
agrees well with theoretical predictions for a model Langevin equation expected to describe fluctuations 
governed by random attachment/detachment of atoms at the step edge \cite{5}. 
This class of behavior is also one of the cases consistent with the previously measured scaling 
of the time correlation function \cite{18}. Since different physical mechanisms will generally 
lead to different values of the persistence exponent, it provides corroborating evidence for 
concluding that the step fluctuations are dominated by attachment/detachment events.

The persistence probability associated with crossing the average step position was both qualitatively 
and quantitatively different from that of the conventional first-return problem. 
There have been few theoretical analyses of the dependence of persistence properties on the 
reference level \cite{19} to guide interpretation of this result. 
However, it is interesting that the time constant of the exponential decay 
for the fixed reference case is found to be independent of large changes in the physical 
parameters of the system. Rescaling the Langevin calculation to account for the change in 
the physical parameters of the experimental system with temperature suggest that this 
observation is physical \cite{20}.

As the number of experimental studies of persistence addressing various theoretical issues increases, 
it will be possible to evaluate the potential applications of the concept. At the least, it provides 
an independent dynamic critical exponent that can be associated with the physical mechanism underlying 
the observed kinetics. As noted by Cueille and Sire \cite{21} for more complicated systems than considered 
in this study, the increased sensitivity of the persistence probability to the details of temporal 
correlations also allows a more complete understanding of the extent to which the usual dynamic critical
exponents uniquely characterize model universality classes. Recent generalizations of persistence 
\cite{21,22,23} may bring added insight to the dynamics of complicated stochastic processes. From an 
applied perspective, given the increased importance of thermal fluctuations with decreasing size 
scale, experimental and theoretical studies of persistence may prove useful in controlling fabrication, 
stability, and response of  nanostructures.

The authors acknowledge useful discussions with T.L. Einstein and  Z. Toroczkai. 
This work was supported by the NSF-MRSEC under DMR-00-80008.


\pagebreak

\begin{figure}
\epsfysize=16.0cm \centerline{\epsffile{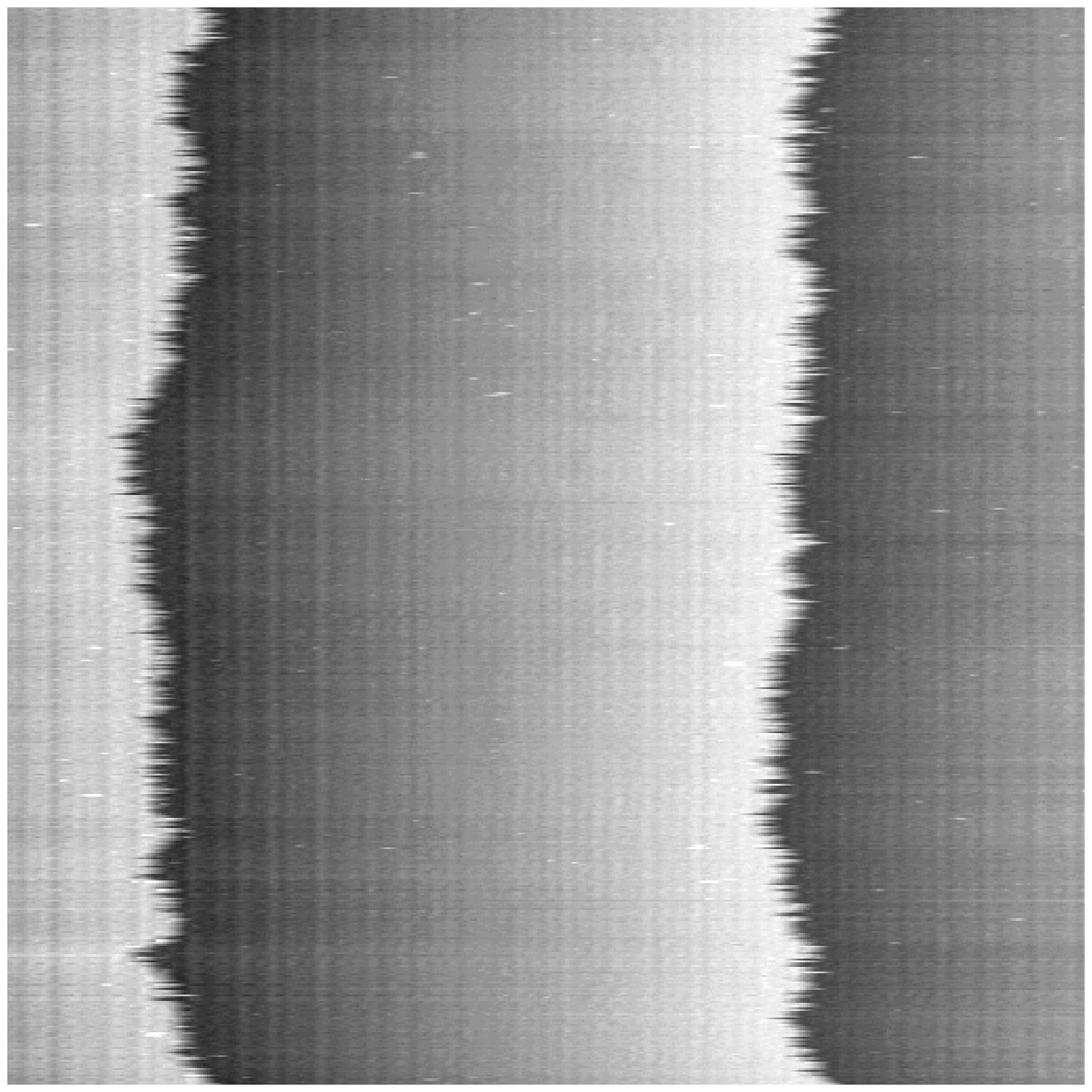}}
\begin{center}
\end{center}
\vspace{0.5 cm}
\caption{Line-scan STM image at temperature $T=970 K$ - repeated measurement at a fixed $y$-coordinate with a 
line-scan length of $100$ nm and a line scan time of $74$ ms. Total time for the measurement is $38$ s.}
\label{figura1}
\end{figure}

\begin{figure}
\epsffile{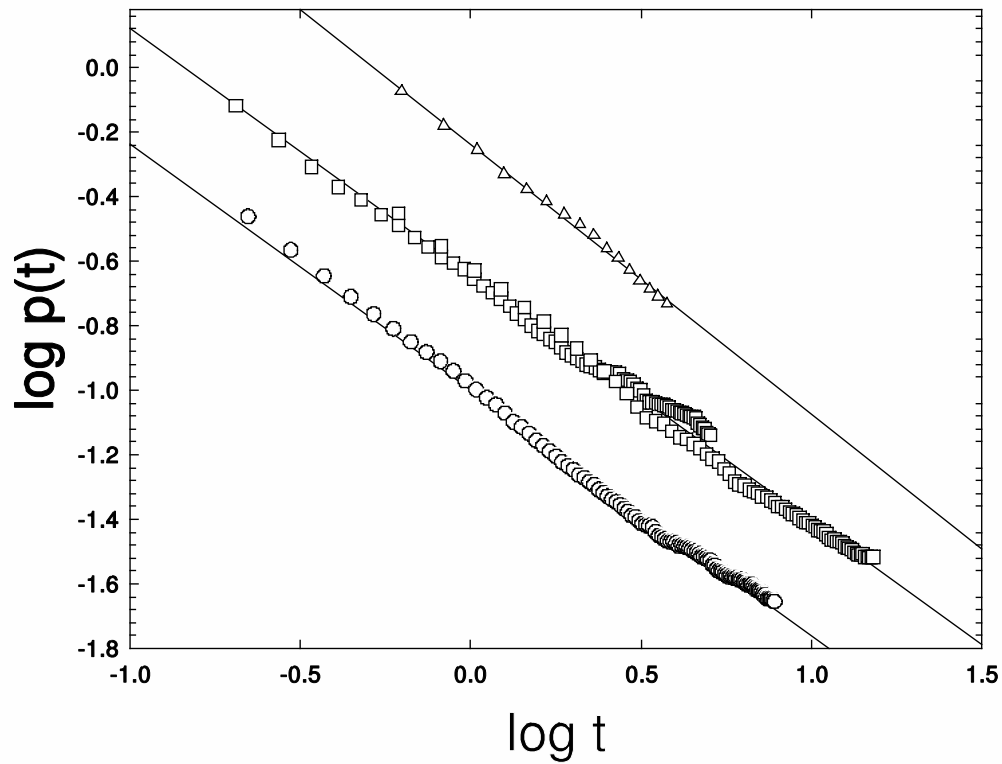}
\epsfysize=8.0cm 
\begin{center}
\end{center}
\vspace{0.5 cm}
\caption {Log-log plot of the measured first-return persistence probabilities for temperatures $T=970K$ (squares), 
$T=870K$ (circles) and $T=770K$ (triangles). The slopes yield values of $\theta=0.77$ for $T=970K$, 
$\theta=0.76$ for $T=870K$, and $\theta=0.84$ for $T=770K$, for an average (weighted by number of steps) 
value of $\theta=0.77 \pm 0.03$.}
\label{figura2}
\end{figure}

\begin{figure}
\epsffile{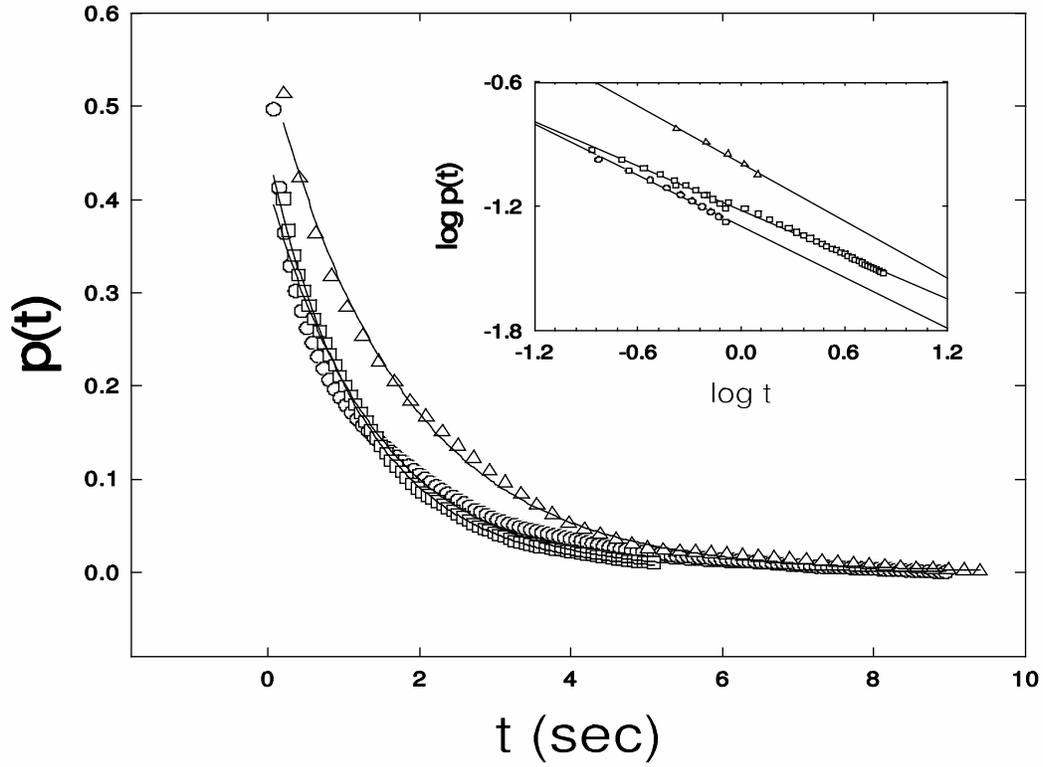}
\epsfysize=8.0cm 
\begin{center}
\end{center}
\vspace{0.5 cm}
\caption {Measured persistence probabilities with respect to the average step position 
fluctuations for three different temperature $T=970K$ (squares), $T=870K$ (circles), 
$T=770K$ (triangles). The solid lines are fits to an exponential decay. The inset shows 
the initial power law scaling of the persistence probabilities which have average aparent 
exponent of $\theta=0.40 \pm 0.03$.}
\label{figura3}
\end{figure}

\begin{figure}
\epsffile{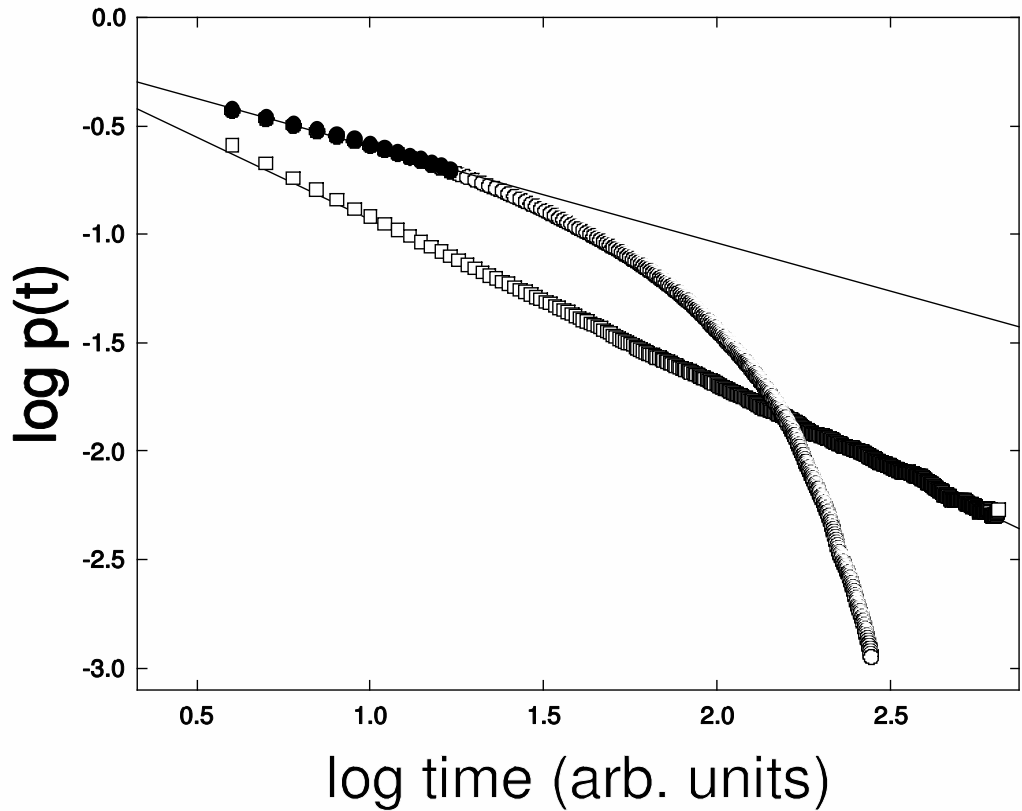}
\epsfysize=8.0cm 
\begin{center}
\end{center}
\vspace{0.5 cm}
\caption {Persistence probabilities obtained from the numerical 
solution of the $\nabla^{2}$ Langevin equation. The slope of the first-return reference 
level calculation (squares) gives a persistence exponent of $0.76$. The average step position 
calculation (circles) shows early time linear behavior with a slope of $0.44$ (solid circles) but at 
later times decays exponentially.}
\label{figura4}
\end{figure}

\end{document}